\documentclass[conference]{IEEEtran}
\IEEEoverridecommandlockouts
\usepackage{cite}
\usepackage{amsmath,amssymb,amsfonts}
\usepackage{algorithmic}
\usepackage{graphicx}
\usepackage{textcomp}
\usepackage{xcolor}

\usepackage{amsmath,epsfig}

\usepackage{adjustbox}
\usepackage{subcaption}
\usepackage{booktabs}
\usepackage{float}
\usepackage{amssymb}
\usepackage{dblfloatfix}
\newcommand{\ra}[1]{\renewcommand{\arraystretch}{#1}}

\begin{document}\sloppy

\newcommand{\linebreakand}{%
      \end{@IEEEauthorhalign}
      \hfill\mbox{}\par
      \mbox{}\hfill\begin{@IEEEauthorhalign}
    }

\title{A Comparison Of Deep Learning MOS Predictors For Speech Synthesis Quality\\
\thanks{This publication has emanated from research conducted with the financial support of Science Foundation Ireland (SFI) under Grant Number \mbox{17/RC-PhD/3483} and 17/RC/2289\_P2. The work of EB was supported by a Turing Fellowship. For the purpose of Open Access, the author has applied a CC BY public copyright licence to any Author Accepted Manuscript version arising from this submission.}
}

\author{\IEEEauthorblockN{Alessandro Ragano}
\IEEEauthorblockA{\textit{School of Computer Science} \\
\textit{University College Dublin}\\
Dublin, Ireland \\
alessandro.ragano@ucd.ie}
\and
\IEEEauthorblockN{Emmanouil Benetos}
\IEEEauthorblockA{\textit{School of EECS} \\
\textit{Queen Mary University of London}\\
London, United Kingdom \\
emmanouil.benetos@qmul.ac.uk}
\and
\IEEEauthorblockN{Michael Chinen}
\IEEEauthorblockA{\textit{Chrome Media Audio} \\
\textit{Google LLC}\\
San Francisco, USA \\
mchinen@google.com}
\and
\IEEEauthorblockN{Helard Becerra Martinez}
\IEEEauthorblockA{\textit{School of Computer Science} \\
\textit{University College Dublin}\\
Dublin, Ireland \\
helard.becerra@ucd.ie}
\and
\IEEEauthorblockN{\phantom{Helard}}
\IEEEauthorblockA{\textit{\phantom{School of}} \\
\textit{\phantom{University}}\\
\phantom{Dublin}\\
\phantom{helard.bece}}
\and
\IEEEauthorblockN{Chandan K A Reddy}
\IEEEauthorblockA{\textit{Chrome Media Audio} \\
\textit{Google LLC}\\
San Francisco, USA \\
chandanka@google.com}
\and
\IEEEauthorblockN{Jan Skoglund}
\IEEEauthorblockA{\textit{Chrome Media Audio} \\
\textit{Google LLC}\\
San Francisco, USA \\
jks@google.com}
\and
\IEEEauthorblockN{Andrew Hines}
\IEEEauthorblockA{\textit{School of Computer Science} \\
\textit{University College Dublin}\\
Dublin, Ireland \\
andrew.hines@ucd.ie}
}

\iftrue
\IEEEpubid{\makebox[\columnwidth]{979-8-3503-4057-0/23/\$31.00
\copyright 2023 IEEE \hfill} \hspace{\columnsep}\makebox[\columnwidth]{ }}
\fi

\maketitle

\begin{abstract}
Speech synthesis quality prediction has made remarkable progress with the development of supervised and self-supervised learning (SSL) MOS predictors but some aspects related to the data are still unclear and require further study.
In this paper, we evaluate several MOS predictors based on wav2vec 2.0 and the NISQA speech quality prediction model to explore the role of the training data, the influence of the system type, and the role of cross-domain features in SSL models. Our evaluation is based on the VoiceMOS challenge dataset.
Results show that SSL-based models show the highest correlation and lowest mean squared error compared to supervised models. The key point of this study is that benchmarking the statistical performance of MOS predictors alone is not sufficient to rank models since potential issues hidden in the data could bias the evaluated performances. 
\end{abstract}
\section{Introduction and Motivations}
\label{sec:1}

The perceived quality of speech synthesis techniques such as text-to-speech (TTS)~\cite{ning2019review} and voice conversion (VC)~\cite{sisman2020overview} is crucial to determine the acceptability of a system. Measuring the quality of synthesized speech is typically carried out with subjective listening tests and objective metrics. Subjective listening tests are the gold standard to assess speech synthesis quality but they are \mbox{time-consuming} and not usable in real-time applications. Objective quality metrics have been proposed to replace listening tests in the above-mentioned scenarios. They are based on predicting the mean opinion score (MOS) of quality without human intervention and are evaluated using quality scores obtained with listening tests. 

Supervised deep learning-based techniques have shown superior performance to traditional objective metrics for both synthetic speech~\cite{patton2016automos,lo2019mosnet,huang2022ldnet} and natural speech (e.g. the NISQA metric~\cite{mittag2021nisqa}). These techniques can be trained directly on the degraded utterances without a reference signal which is beneficial for speech synthesis. 

To mitigate the scarcity of annotated corpora, new speech synthesis MOS prediction techniques based on self-supervised learning (SSL)~\cite{ericsson2022} emerged. Finetuning the model wav2vec 2.0~\cite{baevski2020wav2vec} turns out to generalize better than the other approaches for both utterance and system-level evaluation~\cite{cooper2022generalization}. One advantage of SSL-based techniques like wav2vec~2.0 is to reach high-performance using a relatively small labelled training set.

The above-mentioned techniques have shown superior performance compared to traditional objective metrics. Most of the effort has been put into better architectures but the quality of the data needs further analysis when training speech synthesis MOS predictors to avoid potential data bias. The following properties of the training data are poorly explored and require further analysis and discussion:
\begin{enumerate}
    \item Whether using natural speech data would benefit SSL-based objective MOS predictors for synthetic speech. Since one of the key factors of high-quality synthetic speech is naturalness~\cite{nusbaum1997measuring}, augmenting the synthetic fine-tuning set with natural speech observations that are labeled with MOS could improve model performance. Deep models typically benefit from high diversity in the training dataset. However, training MOS predictors with synthesized utterances is limited by the availability of synthesis systems that are used. Natural speech instead has the advantage of simulating artificial degradations to help increase dataset diversity. 
    
    \item Whether objective MOS predictors perform similarly on both VC and TTS synthesis types. Training sets might include both VC and TTS synthesized utterances where each exhibits unique acoustic properties. Exploring how the models perform on the synthesis types is important since raters tend to prefer TTS over VC utterances~\cite{cooper2021voices}. The risk of introducing a confounding factor in the training set is high if datasets are designed with low MOS utterances sampled from VC systems and high MOS utterances sampled from TTS systems. In the latter case, objective metrics might learn to discriminate the two system types instead of learning the quality prediction task.
    \item Whether finetuning wav2vec~2.0 can be further improved by combining different feature representations. The usage of cross-domain features from both frequency and time-domain representations has not been explored yet for speech synthesis quality. The model wav2vec 2.0 is designed to learn feature representations that are sufficient by themselves, avoiding the need for handcrafted features such as mel spectrograms. However, it is not clear whether the combined approach can further improve the quality prediction performance.
\end{enumerate}

In this paper, we present a comparison of speech synthesis MOS predictors addressing the above-mentioned issues related to the data. 
The experiments in this paper are all conducted using the VoiceMOS dataset~\cite{cooper2022generalization} for synthetic speech, the NISQA~\cite{mittag2021nisqa} and the PSTN~\cite{mittag2020dnn} corpus for natural speech. Assessing these issues will assist with curating better datasets for speech synthesis quality prediction and understanding better the limits of the state-of-the-art architectures. 

\begin{table}[b]
\caption{An overview of the evaluated model. The models Fusion 1 and Fusion 2 are pretrained with Librispeech960 (LS960) for wav2vec 2.0 solving the SSL task while Librispeech100 (LS100) is used to pretrain the framewise CNN by training an autoencoder (AE).}
\centering
\Huge
\ra{0.8}
\begin{adjustbox}{max width=0.49\textwidth}
\begin{tabular}{@{}lclllllllllll@{}}\toprule
\multicolumn{1}{c} {} & \multicolumn{1}{c}{}\\
Model && Architecture && Train Set && Pretrain Dataset && Pretrain Task\\ \midrule
ConvMaxPool && ConvMaxPool && VoiceMOS Train  && // && // \\
ConvMaxPool* && ConvMaxPool  && VoiceMOS Train && LS100 && AE\\
NISQA && NISQA  && VoiceMOS Train && // && // \\
w2v\_VoiceMOS && w2vMOS  && VoiceMOS Train  && LS960 && SSL \\
w2v\_NISQA && w2vMOS  && NISQA corpus && LS960 && SSL \\
w2v\_PSTN && w2vMOS && PSTN corpus && LS960 && SSL \\
Fusion 1 && Fusion 1  && VoiceMOS Train && LS960, LS100 && SSL, AE \\
Fusion 2 && Fusion 2 && VoiceMOS Train && LS960, LS100 && SSL, AE \\

\bottomrule
\label{tab:models}
\end{tabular}
\end{adjustbox}
\end{table}

\section{Architecture Comparison}
In this section, we describe all the architectures that we evaluate. Throughout the paper, we distinguish the term model from architecture. We use the term model to refer to an architecture trained on a specific dataset. For example, wav2vec 2.0 trained on the VoiceMOS dataset is a different model from wav2vec 2.0 trained on the NISQA corpus but they have the same architecture. Using this terminology, we first describe the architectures that are used and then the models by specifying on which datasets the architectures are trained. A detailed overview of the architectures and models used in this paper is reported in Table \ref{tab:models}.

\subsection{NISQA}
NISQA has been proposed to predict natural speech quality and it is based on a framewise CNN that maps a mel spectrogram patch of size $48\times15$ ($48$ mel bands, $15$ time frames)  to $6\times1\times64$ features where $64$ is the number of filters of the last convolutional layer. The features are flattened into $384$-dimensional vectors and fed to a self-attention layer. The latter creates feature vectors that capture the time dependency of the patches across the whole speech utterance. Self-attention output vectors are the input of an attention-pooling network that predicts a single continuous value (MOS).

\subsection{ConvMaxPool}
The NISQA architecture is made of $\approx218K$ parameters. To evaluate a lighter network we apply the following changes to the NISQA architecture. At the output of the framewise CNN, we extract features using a global average pooling (GAP) layer. The latter averages all the feature maps of the last layer producing $64$-dimensional feature vectors, one corresponding to each spectrogram patch. The dimension of the GAP layer output depends on the number of filters used in the last convolutional layer which is $64$ in our case. 
Next, we apply temporal max-pooling on the feature sequence obtaining one $64$-dimensional vector representing the whole speech utterance. Finally, a fully connected (FC) layer is used to predict MOS. To predict MOS within the range 1-5 we apply a shifted and scale sigmoid at the output of the FC layer. 
These changes give us an architecture 40\% smaller than NISQA the number of parameters ($\approx$135K vs. $\approx$218K).
This architecture is called ConvMaxPool which stands for \underline{conv}olutional \underline{max}-\underline{pool}ing.

Pretraining using deep autoencoders has been effective for speech~\cite{ragano2021more,soni2016novel} and audio-visual quality prediction~\cite{martinez2019navidad}. For this reason, we pretrain the framewise CNN block with an autoencoder trained on a partition of the Librispeech corpus\footnote{To pretrain the autoencoder we use the Librispeech partition 100-train-clean that we call Librispeech100.}~\cite{panayotov2015librispeech}. The model pre-trained with the autoencoder is called ConvMaxPool*.

\begin{figure}[htb]
	\begin{minipage}[b]{1.0\linewidth}
		\centering
		\centerline{\includegraphics[width=0.80\textwidth]{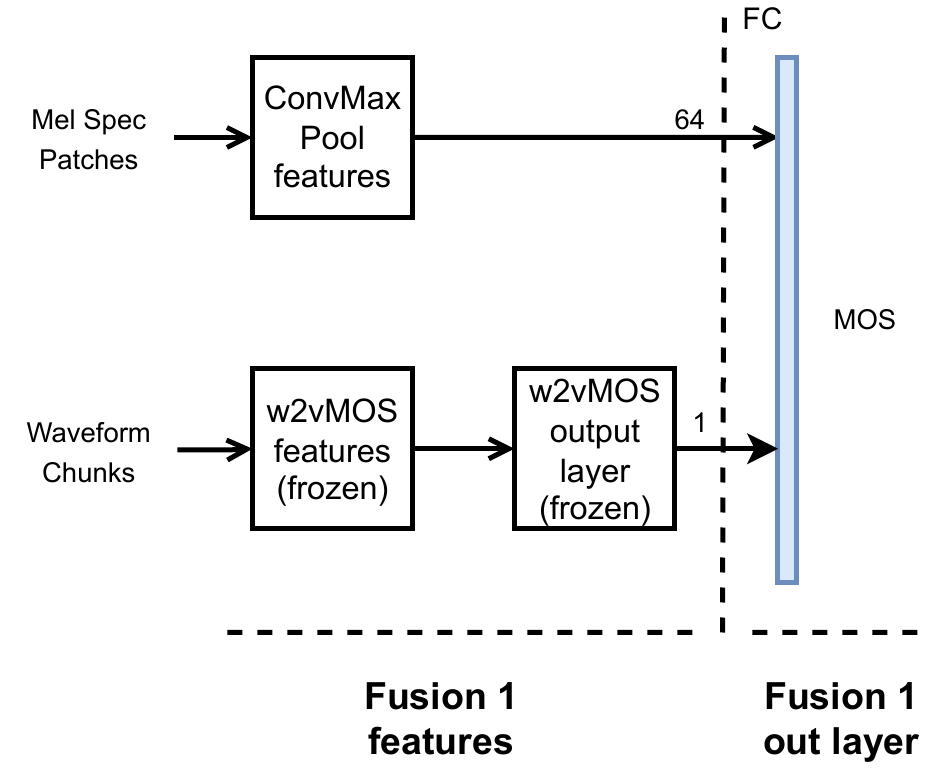}}
	\end{minipage}
	\caption{Fusion 1. This architecture combines the ConvMaxPool features with the predicted MOS of the base version of w2vMOS. During training, the weights of the ConvMaxPool features and the FC layer are optimized while the rest is not updated.}
	\label{fig:fusion1}
\end{figure}
\begin{figure}[!htb]
	\begin{minipage}[b]{1.0\linewidth}
		\centering
		\centerline{\includegraphics[width=0.80\textwidth]{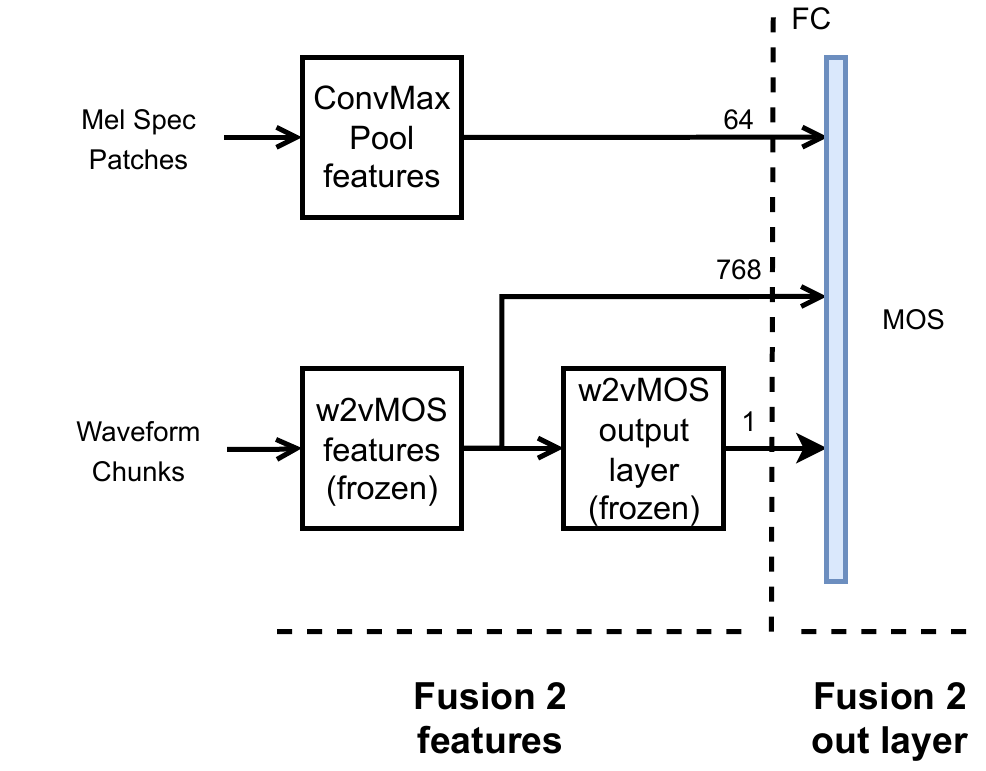}}
	\end{minipage}
	\caption{Fusion 2. This architecture combines the ConvMaxPool features with the predicted MOS and the features of the base version of w2vMOS. During training the weights of ConvMaxPool features and the FC layer are optimized while the rest is not updated.}
	\label{fig:fusion2}
\end{figure}

\subsection{SSL Models}
Finetuning the SSL model wav2vec 2.0 showed the best performance in the VoiceMOS challenge 2022~\cite{cooper2022generalization}. We use only the BASE version which is composed of 12 Transformer layers and 768-dimensional feature vectors for each time frame. This architecture is called w2v\_MOS.

Both ConvMaxPool and NISQA use mel spectrograms as input features while w2vMOS works directly from the time-domain waveform. To understand whether wav2vec 2.0 finetuned features show benefit if fused with spectral domain features, we evaluate 2 fusion approaches combining ConvMaxPool and w2vMOS.

The first approach \textit{Fusion 1} combines ConvMaxPool features with the predicted MOS of w2vMOS (Figure \ref{fig:fusion1}). The second approach \textit{Fusion 2} further adds the features from w2vMOS (Figure \ref{fig:fusion2}). It should be noted that we add features from w2vMOS which are the weights that are finetuned on the MOS prediction task and not the ones trained to solve the SSL task. 

\section{Model Comparison}
This section describes all the models that are used. The term model refers to an architecture trained on a specific dataset. To describe the models we first outline a brief overview of the datasets used.

\subsection{Datasets}
The VoiceMOS dataset~\cite{cooper2022generalization} represents the corpus employed for the VoiceMOS challenge. It is divided into the main track and the out-of-domain track. Only the main-track data is used in this study. The main-track corpus is composed of 187 speech synthesis systems including TTS, VC, and natural speech. The speech synthesis systems include TTS utterances from the Blizzard challenge (BC) and the ESPNet systems, while the VC utterances are taken from the Voice Conversion Challenge (VCC). Natural speech is taken from all the 3 system types (BC, VCC, and ESPNet).
Several speakers and utterances are collected and a listening test has been conducted in the lab using 8 listeners for each speech sample~\cite{cooper2021voices}.
The VoiceMOS dataset is split into training, validation, and test set. The training set of the VoiceMOS corpus includes roughly 5k training samples.

The NISQA corpus has been used to design the NISQA architecture. It is composed of 14k natural speech samples degraded with both real-world conditions (e.g. video call services and mobile phones) and synthetic degradations (e.g. codecs, background noise). From the NISQA corpus, we only use the training sets. 5 listeners are used on average for rating the speech samples of the training sets which account for roughly 11k samples. 

The PSTN corpus includes $\approx$80k speech samples with public switched telephone network degradations obtained with real phone calls. 
Due to the large size of the PSTN corpus, we extract a random subset of 7k samples that has the same distribution as the original size.

\subsection{Models}
All the architectures are trained using the VoiceMOS training partition while the architecture w2vMOS is also trained on the NISQA and the PSTN corpora. The architecture ConvMaxPool is evaluated both with the autoencoder and trained from scratch. 
To explore the augmentation of the synthetic speech dataset with natural speech, a model fine-tuned on the combination of the two dataset types should be trained. However, we decided to avoid this approach since the label space between two different datasets is not the same. For instance, the NISQA corpus and the VoiceMOS dataset have been labeled with two different methodologies and different content stimuli. This implies that each dataset has some unique bias and that combining datasets will invalidate the ranking of the MOS labels. Instead of combining the datasets, we measure how using natural speech only affects the performance of synthetic speech to gain insight into the utility of the natural speech data. In the case promising results are obtained, a suggestion would be to add natural speech when labelling synthetic speech or to combine dataset by learning the bias in each corpus which is outside the scope of this paper.

\subsection{Training}
The models ConvMaxPool, ConvMaxPool*, NISQA, Fusion 1, and Fusion 2 are trained with Adam optimizer and a learning rate of 0.001 while the models using the architecture w2vMOS (w2v\_VoiceMOS, w2v\_NISQA, and w2v\_PSTN) are trained with stochastic gradient descent using 0.0001 as the learning rate.
Training is stopped if the loss measured on the validation set did not improve after 20 epochs. The L1 loss is used as done in \cite{cooper2022generalization} and it is calculated per speech sample during training. The validation set used for the early stopping of training is always the VoiceMOS validation partition because we compare all the performances on the VoiceMOS test set. Choosing similar validation and test sets is important to avoid misleading results on the test set. 

\section{Results}
The models are all compared on the VoiceMOS test set using the mean squared error (MSE), the linear correlation coefficient (LCC), and the Spearman's rank correlation coefficient (SRCC) which are common measures to evaluate objective speech synthesis quality metrics~\cite{cooper2022generalization}.
Performance measures are calculated per utterance and per system where the predictions and the ground truth are aggregated by taking the mean of all the speech samples synthesized with the same speech synthesis system. Each model is trained 10 times with random weight initialization for the non-pretrained layers since initial experiments were showing some differences of the combined models (Fusion 1, Fusion 2) compared to w2v\_VoiceMOS depending on the weight initialization.

The compared models have been trained with corpora that are labeled with different subjective experiments that do not have common anchors. This affects the MSE performance since MOS is mapped to a different scale in each training set. To compensate for the variance of each listening test, we fit a first-degree polynomial of the predicted MOS with the ground truth MOS of the VoiceMOS test set as recommended by the ITU-T Rec. P.1401~\cite{itu-1401}.

Table \ref{tab:test_res} shows the model performance using the mean and the 95\% confidence intervals over the 10 different runs. The results in Table \ref{tab:test_res} show that all the models based on wav2vec outperform the others.
To interpret the results from Table \ref{tab:test_res} we conduct a statistical analysis based on the analysis of variance (ANOVA) and post-hoc corrections for both utterance-level and system-level results. The analysis is needed to avoid misleading conclusions which might depend on a particular model instance. 
The multiple comparison analysis is conducted with the Tukey method after obtaining that the null hypothesis (i.e. all the means are equal) is rejected with ANOVA.
\begin{table*}[!t]
\caption[C]{Test Set Results.}
\centering
\Huge
\ra{0.8}
\begin{adjustbox}{max width=0.70\textwidth}
\begin{tabular}{@{}lclccclclclccccccll@{}}\toprule

& \multicolumn{6}{c} {\textbf{Utterance Level}} & \multicolumn{6}{c}{\textbf{System Level}}\\
\cmidrule{2-6} \cmidrule{8-12}

&  MSE && LCC && SRCC && MSE && LCC && SRCC &  \\ \midrule
ConvMaxPool & 0.36$\pm$0.01 && 0.76$\pm$0.01 && 0.76$\pm$0.01 && 0.19$\pm$0.01 && 0.85$\pm$0.01 && 0.85$\pm$0.01  \\
ConvMaxPool* & 0.31$\pm$0.01 && 0.80$\pm$0.01 && 0.79$\pm$0.01 && 0.16$\pm$0.01 && 0.87$\pm$0.01 && 0.87$\pm$0.01\\
NISQA & 0.30$\pm$0.01 && 0.80$\pm$0.01 && 0.80$\pm$0.01 && 0.14$\pm$0.01 && 0.89$\pm$0.01 && 0.89$\pm$0.01 \\
w2v\_VoiceMOS & \textbf{0.20$\pm$0.01} && 0.87$\pm$0.00 && 0.87$\pm$0.00 && \textbf{0.10$\pm$0.01} && \textbf{0.93$\pm$0.01} && \textbf{0.92$\pm$0.01} \\
w2v\_PSTN & 0.42$\pm$0.02 && 0.71$\pm$0.02 && 0.75$\pm$0.02 && 0.27$\pm$0.02 && 0.78$\pm$0.01 && 0.80$\pm$0.02 \\
w2v\_NISQA & 0.34$\pm$0.02 && 0.78$\pm$0.01 && 0.77$\pm$0.01 && 0.20$\pm$0.02 && 0.84$\pm$0.01 && 0.84$\pm$0.01 \\
Fusion 1 & \textbf{0.20$\pm$0.00} && \textbf{0.88$\pm$0.01} && 0.87$\pm$0.00 && \textbf{0.10$\pm$0.00} && \textbf{0.93$\pm$0.00} && \textbf{0.92$\pm$0.00} \\
Fusion 2 & \textbf{0.20$\pm$0.00} && \textbf{0.88$\pm$0.00} && \textbf{0.88$\pm$0.00} && \textbf{0.10$\pm$0.00} && 0.92$\pm$0.00 && \textbf{0.92$\pm$0.00} \\
\bottomrule
\label{tab:test_res}
\end{tabular}
\end{adjustbox}
\end{table*}

\begin{table}[!t]
\caption[C]{Post-hoc Tukey, \checkmark means that the null hypothesis (i.e. equal means) is not rejected (p-value$\geq$0.05). Only the pairs with at least one not rejected null hypothesis are reported.}
\centering
\Huge
\ra{0.8}
\begin{adjustbox}{max width=0.48\textwidth}
\begin{tabular}{@{}lclccclclclccccccll@{}}\toprule

& \multicolumn{6}{c} {\textbf{Utterance Level}} & \multicolumn{6}{c}{\textbf{System Level}}\\
\cmidrule{2-6} \cmidrule{8-12}

&  MSE && LCC && SRCC && MSE && LCC && SRCC &  \\ \midrule
ConvMaxPool*-NISQA          & \checkmark && \checkmark && \checkmark && \checkmark && \checkmark && \checkmark \\
Fusion1-Fusion2             & \checkmark && \checkmark && \checkmark && \checkmark && \checkmark && \checkmark \\
Fusion1-w2v\_VoiceMOS       & \checkmark && \checkmark && \checkmark && \checkmark && \checkmark && \checkmark \\
Fusion2-w2v\_VoiceMOS       & \checkmark && \checkmark && \checkmark && \checkmark && \checkmark && \checkmark \\
ConvMaxPool-w2v\_NISQA      &  \checkmark && \checkmark && \checkmark && \checkmark && \checkmark && \checkmark \\
ConvMaxPool-w2v\_PSTN       &  && && \checkmark && && && \\
ConvMaxPool*-w2v\_NISQA     &  && && \checkmark && && && \\

\bottomrule
\label{tab:post-hoc}
\end{tabular}
\end{adjustbox}
\end{table}

\begin{figure}[!t]
\centering
	\begin{minipage}[b]{0.40\textwidth}
		\centering
		\centerline{\includegraphics[width=\textwidth]{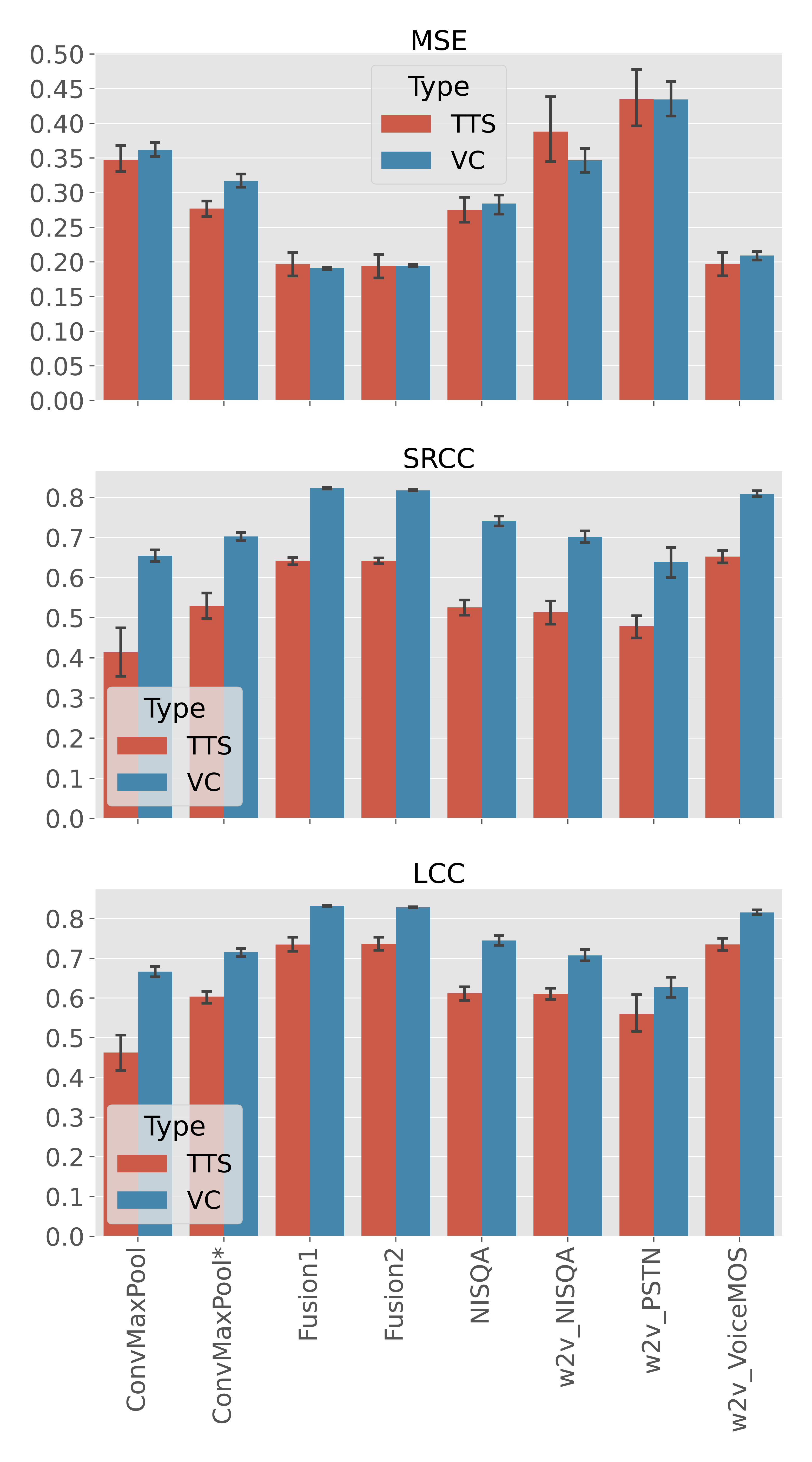}}
	\end{minipage}
	\caption{Utterance-level performance comparison (MSE, SRCC, LCC) between TTS (BC, ESPNet) and VC (VCC) subsets.}
	\label{fig:sys_type}
    \end{figure}

The comparison analysis reported in Table \ref{tab:post-hoc} shows that NISQA and ConvMaxPool* do not show any difference as well as both Fusion models and w2v\_VoiceMOS. Likewise, ConvMaxPool and w2v\_NISQA show no difference in all the performance metrics. We only report the comparisons where at least one null hypothesis is not rejected by the Tukey test. Therefore, all the pairs not reported in Table \ref{tab:post-hoc} show statistically different performances in all the measures.   

\subsection{Performance by System Type}
The results above do not inform whether the performance difference is related to a particular synthesis system type. The VoiceMOS dataset includes text-to-speech utterances from BC and ESPNet, voice conversion samples from VCC systems, and natural speech. 
To conduct this investigation, we use utterance-level performances since system-level metrics could be less informative due to the non consistent sample size (i.e. the number of utterances per each system) between the systems in the VoiceMOS test set~\cite{chinen2022using}. Indeed, we found that 4 out of 5 systems with poorest system-level performances all belong to TTS and unlike VC systems all exhibit a low sample size. Therefore, system-level performances might mislead the comparison between TTS and VC speech.
Figure \ref{fig:sys_type} shows the mean and the 95\% confidence intervals of MSE, SRCC, and LCC. It can be seen that correlations are higher for VC utterances in all the models but the MSE is the same in all the models except for ConvMaxPool*.

\section{Discussion}
In Section 1 we asked whether SSL-based models for synthetic speech MOS prediction can benefit from the usage of natural speech data. 
The results show that the wav2vec 2.0 architecture is sensitive to the training set used. Although finetuning wav2vec 2.0 on natural speech (w2v\_NISQA and w2v\_PSTN) shows comparable performance with other architectures trained on synthetic speech (ConvMaxPool) they are worse than w2v\_VoiceMOS which is finetuned on synthetic speech. So, using natural speech datasets like the NISQA and the PSTN corpora to finetune wav2vec 2.0 does not generalize well on synthetic speech quality. This implies that they might be less useful for augmenting the synthetic speech training set. 
We suggest that other confounding factors beyond natural speech could also contribute to the domain gap and require further discussion. 

For example, the MOS distribution of the NISQA Corpus training set (Figure \ref{fig:mos_dis} (a)) shows a lower resolution than the VoiceMOS training set (Figure \ref{fig:mos_dis} (b)). Unlike the VoiceMOS set which is labeled in the lab, the NISQA Corpus training set is annotated with crowdsourcing using 5 listeners on average. The discrimination between close signals in quality is likely lost in the training set of the NISQA Corpus.
Another interpretation of the domain gap between natural and synthetic speech data might be related to the overlap between the speech synthesis systems in the training and the test set of the VoiceMOS dataset. Chinen et al.~\cite{chinen2022using} showed that using the system ID as the only input feature contains a reasonable amount of information on the quality scores in the VoiceMOS set. This might suggest that the model trained on the VoiceMOS training set (w2v\_VoiceMOS) might learn to discriminate the synthesis systems which represents a confounding factor for quality. 
This study found no improvement when training with natural speech but we suggest that the overlap of the systems in the in-domain data scenario (i.e. same synthetic speech systems in both training and test data) should be explored in more detail to better understand if the performance evaluations are fair.

\begin{figure}[t]
	\begin{minipage}[b]{0.24\textwidth}
		\centering
		\centerline{\includegraphics[width=\textwidth]{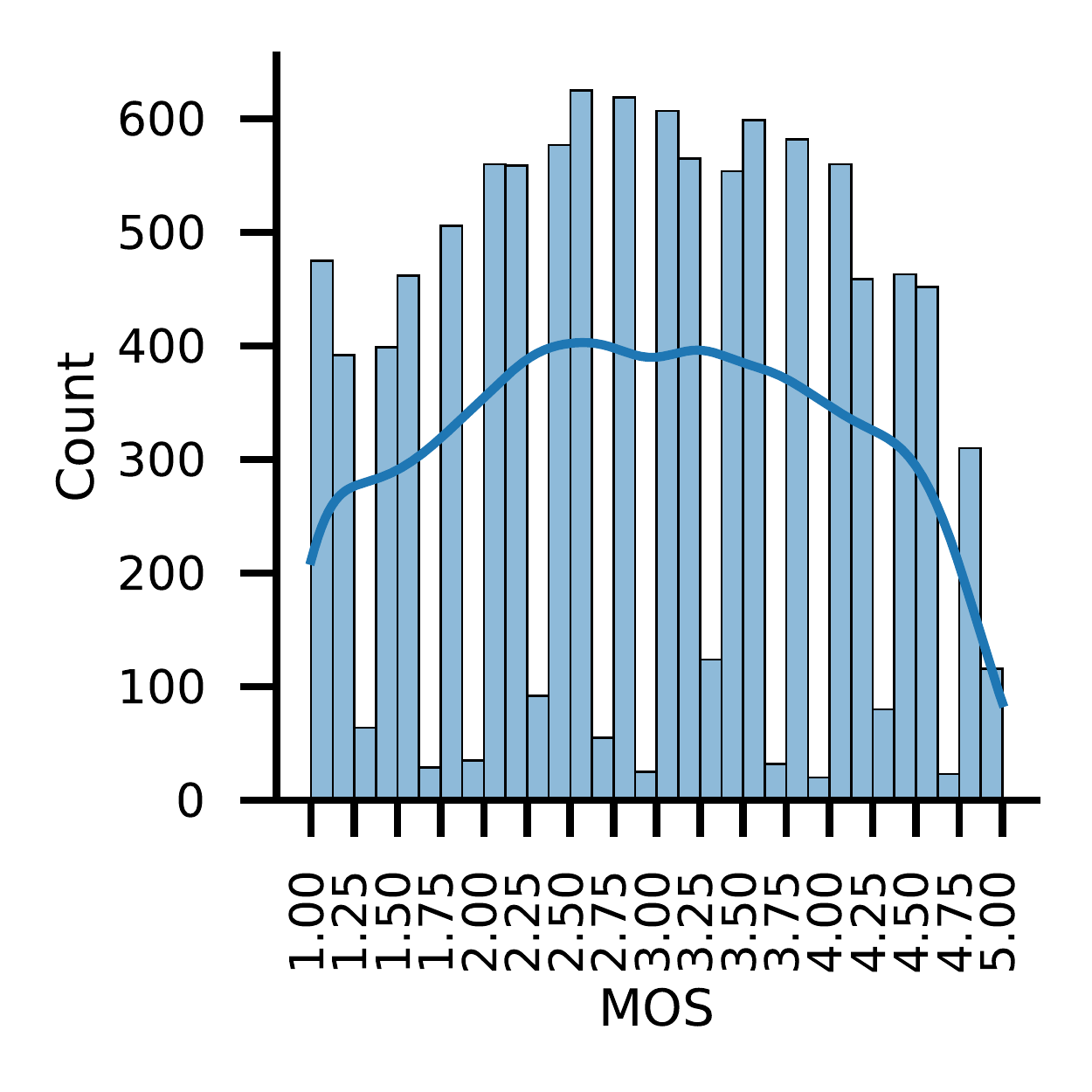}}
		\centerline{(a) NISQA Corpus}\medskip
	\end{minipage}
	\begin{minipage}[b]{0.24\textwidth}
		\centering
		\centerline{\includegraphics[width=\textwidth]{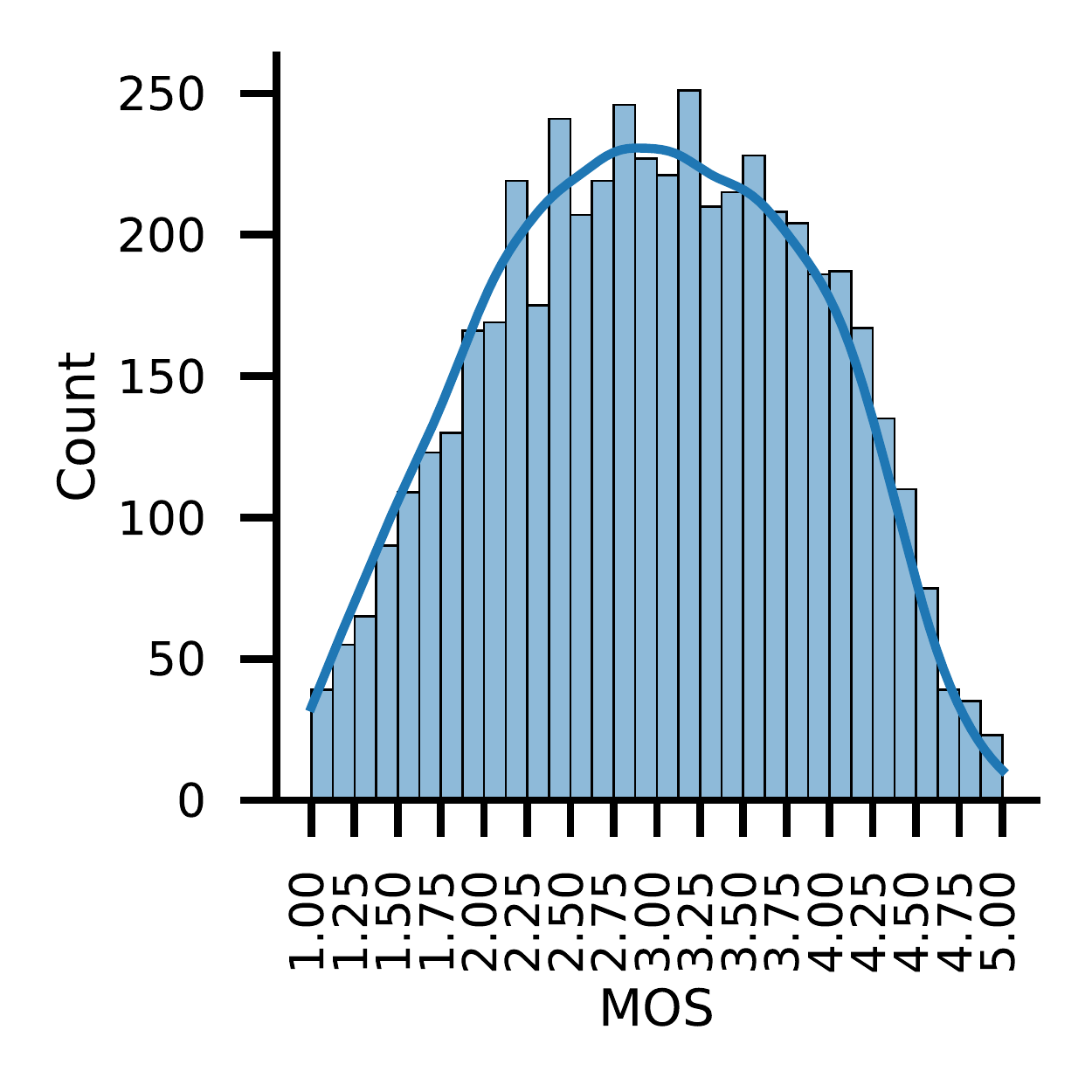}}
		\centerline{(b) VoiceMOS}\medskip
	\end{minipage}
	\caption{MOS histogram of the NISQA Corpus training set (a) and the VoiceMOS training set (b).}
	\label{fig:mos_dis}
\end{figure}

The predictors evaluated in this paper do not exhibit similar quality prediction performance for TTS utterances (BC, ESPNet) and VC utterances (VCC).
Despite the number of VC utterances is smaller than the TTS ones (ESPNet=5.37\%, Natural=5.45\%, VCC=37.41\%, BC=51.77\%) the correlation scores are higher for voice conversion utterances in all the models while the MSE is the same. The contrast between SRCC and MSE means that both monotonic (SRCC) and linear (LCC) relationship are weaker in the case of TTS utterances despite the same MSE. We suggest that this could be explained by how the MOS labels of VC and TTS systems are distributed in the dataset. Figure~\ref{fig:mos_dis_systype} shows the MOS histogram of TTS systems (BC, ESPNet), VC systems (VCC), and natural speech in both the training and test set of the VoiceMOS corpus where 5 histogram bins are obtained from the MOS percentile calculated per-system. It can be seen how the lowest MOS group consists of mostly VC systems while the highest MOS group is heterogeneous and includes all the system types.
This implies that the model might learn to map some acoustic properties that are unique of VC systems to bad quality.
More specifically, our analysis suggests that mixing together VC systems and TTS systems affects the model capacity to predict a monotonic or linear relationship of TTS techniques. Speech synthesis based on TTS has advanced significantly in recent years (e.g. ESPNet systems) which encourages the development of MOS predictors that can discriminate the nuances between high-quality TTS systems. This opens the question as to whether datasets that exhibit a higher representations of recent TTS systems and no presence of VC systems might be more advantageous to develop MOS predictors for TTS.

\begin{figure}[t]
\centering
	\begin{minipage}[b]{0.38\textwidth}
		\centering
		\centerline{\includegraphics[width=\textwidth]{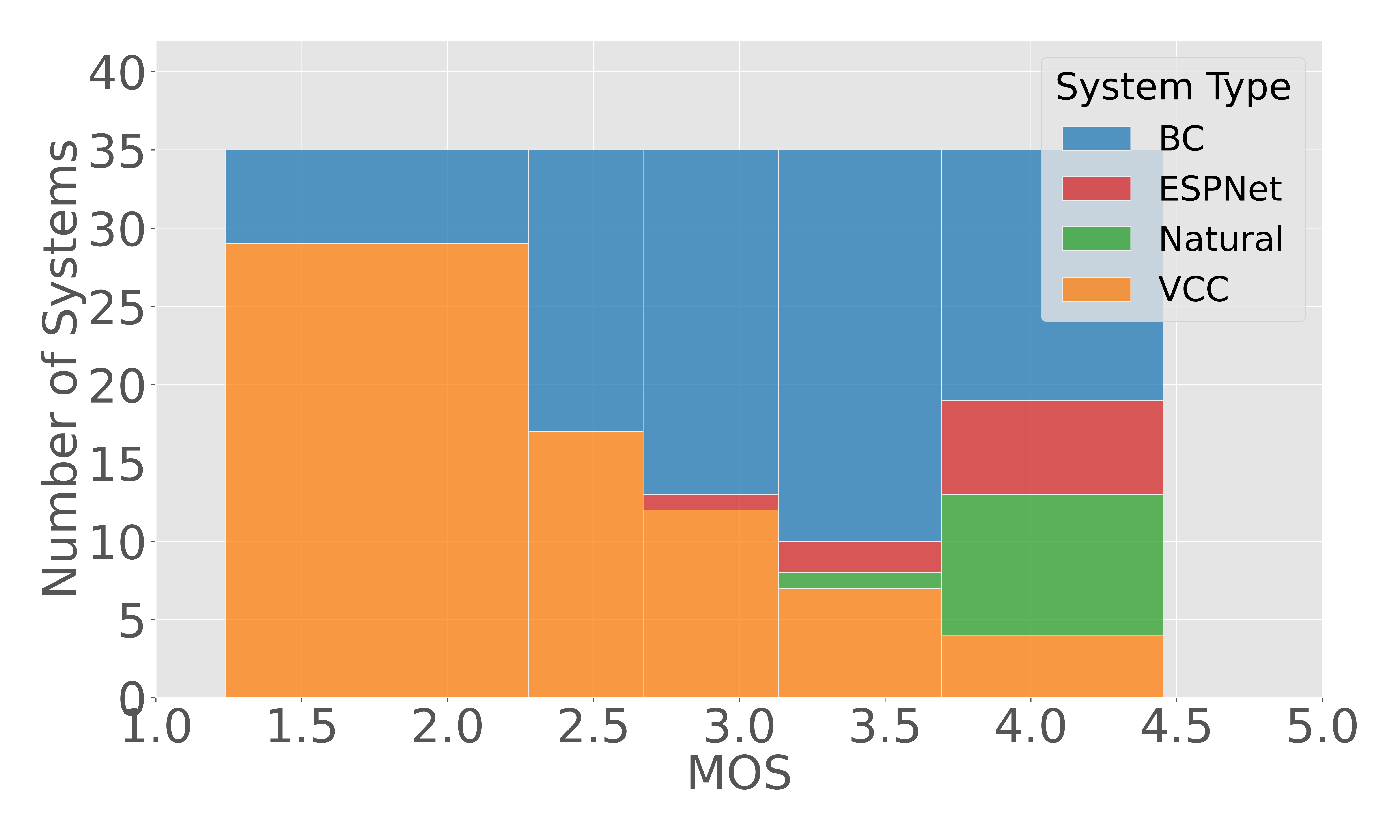}}
		\centerline{(a) VoiceMOS training set}\medskip
	\end{minipage}
	\begin{minipage}[b]{0.38\textwidth}
		\centering
		\centerline{\includegraphics[width=\textwidth]{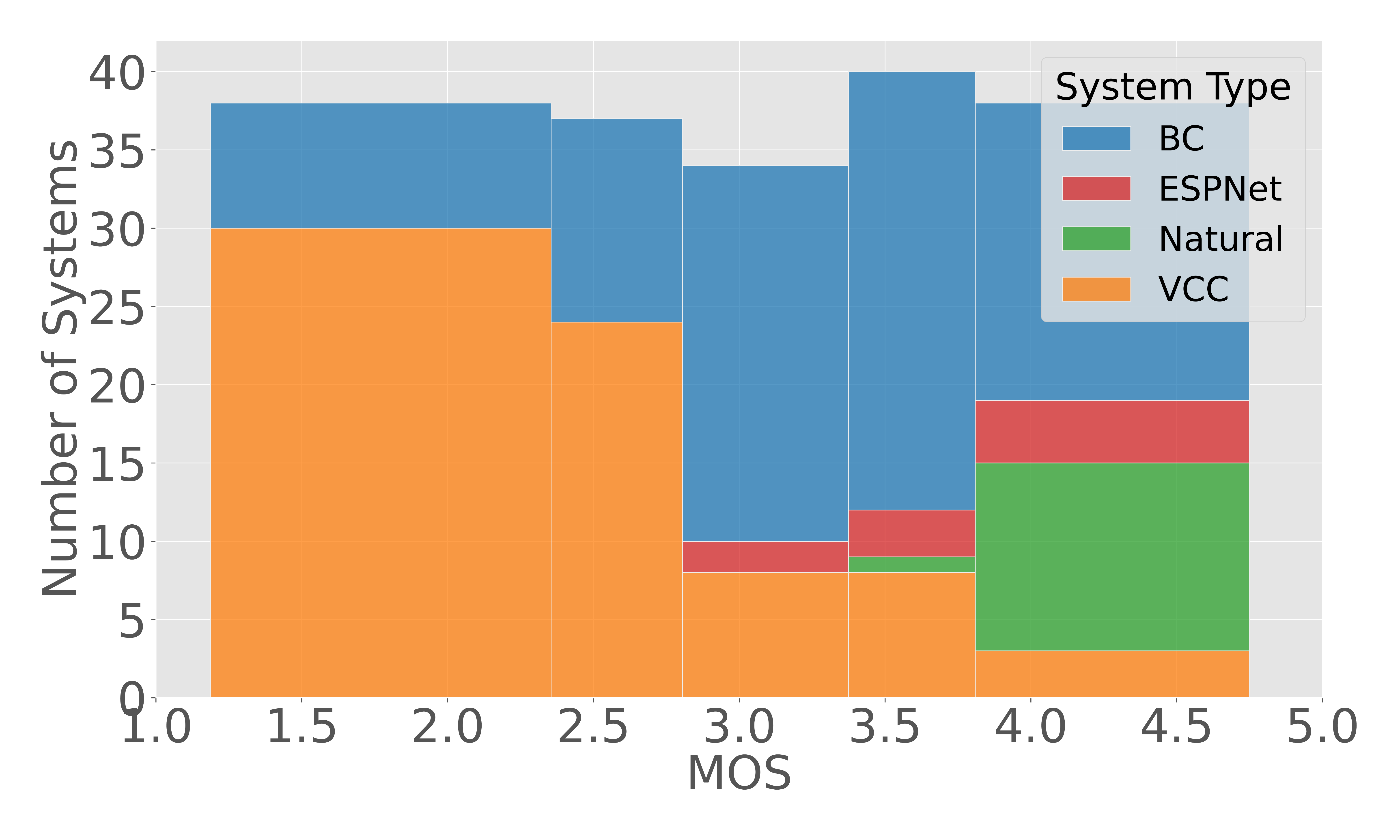}}
		\centerline{(b) VoiceMOS test set}\medskip
	\end{minipage}
	\caption{MOS histogram with 5 bins obtained from the MOS percentile of TTS (BC, ESPNet), VC (VCC), and natural synthesis systems in the VoiceMOS training set (a) and the VoiceMOS test set (b). MOS is calculated per-system.}
	\label{fig:mos_dis_systype}
    \end{figure}

We also posed a question in Section 1 concerning the combination of time and spectral feature representations of the data. The multiple comparisons showed that the two fusion approaches (Fusion 1, and Fusion 2) do not improve w2vMOS. Even though some small differences occur between the fusion-based approaches and w2v\_VoiceMOS, overall the statistical analysis does not show any difference. This result confirms that wav2vec 2.0 learns robust features from the raw waveform which allows us to avoid hand-crafted features like mel spectrograms. Finally, we showed that the performance difference between NISQA and ConvMaxPool* is not statistically significant. ConvMaxPool* allows us a 40\% reduction of the network size with the expense of training an autoencoder first. However, having a lighter network is advantageous when memory constraints occur.

\section{Conclusions}
In this paper, we have compared several speech synthesis MOS predictors based on wav2vec 2.0 and NISQA to explore some potential problems in the dataset used. Our experiments highlight the following issues in the data. 1) SSL-based models generalize less when trained on on out-of-domain data (natural speech labeled from different listening test). 2) The system types play a key role in the VoiceMOS training set as we found that all the evaluated models predict the quality of voice conversion systems consistently better than text-to-speech systems (SRCC, LCC) despite the VoiceMOS training set is imbalanced in favor of TTS audio. 3) Combined feature representations (raw audio and mel spectrograms) does not affect the correlation with listening test MOS confirming the ability of pre-trained SSL-based models to learn robust features from raw audio.

\bibliographystyle{IEEEbib}
\bibliography{refs}
\end{document}